# Bound states in the continuum in fiber Bragg gratings

Xingwei Gao,[1,2] Bo Zhen,[3] Marin Soljačić,[4] Hongsheng Chen,[1,5,*] and Chia Wei Hsu[6,*]

[1]State Key Laboratory of Modern Optical Instrumentation, College of Information Science & Electronic Engineering, Zhejiang University, Hangzhou 310027, China
[2]Department of Electrical and System Engineering, University of Pennsylvania, Philadelphia, Pennsylvania 19104, USA
[3]Department of Physics and Astronomy, University of Pennsylvania, Philadelphia, Pennsylvania 19104, USA
[4]Department of Physics, Massachusetts Institute of Technology, Cambridge, Massachusetts 02139, USA
[5]The Electromagnetics Academy at Zhejiang University, Zhejiang University, Hangzhou 310027, China
[6]Ming Hsieh Department of Electrical Engineering, University of Southern California, Los Angeles, California 90089, USA
*Correspondence authors: C.W.H. (Email: cwhsu@usc.edu) and H.C. (Email: hansomchen@zju.edu.cn)

**Optical fibers typically confine light through total internal reflection or through photonic bandgaps. Here we show that light can be perfectly guided in optical fibers through a different mechanism based on bound states in the continuum (BICs). In fibers with periodic Bragg gratings, we predict bona fide BICs in pure-polarization modes, as well as quasi-BICs in hybrid-polarization modes. These guided modes exist robustly without the need for fine structural tuning, and they persist even with the very small grating index contrasts that are available in conventional fiber Bragg gratings. The suppression of radiation loss arises from the coupling between a weakly-radiating mode and a strongly-radiating one. This finding opens the possibility of guiding light with BICs in optical fibers and their applications in distributed fiber sensors, in-line fiber filters, and high-power fiber lasers.**

## 1. INTRODUCTION

The ability to guide light within optical fibers is important for basic sciences as well as for a wide range of applications, from nonlinear optics [1] and fiber lasers [2-5] to telecommunications [6, 7], sensors [8, 9], imaging [4, 10], and surgery [11], to name only a few. Typical optical fibers confine light through total internal reflection or via photonic bandgaps [12-14]; in such fibers, light cannot escape because only evanescent fields are allowed outside the fiber core. In recent years, a new paradigm for confining light emerged based on the concepts of bound states in the continuum (BICs) [15] that originated in quantum mechanics [15-17]. A BIC can achieve perfect confinement even in the presence of radiating modes in the free space. While conventional guided modes exist across a continuous range of frequencies and propagation constants, a BIC generally exists only at an isolated frequency at an isolated propagation constant [18], therefore providing much more modal selectivity. Optical BICs received much interest and have been experimentally realized in a variety of systems [15, 18-28]. If realized in optical fibers, BICs can have many promising uses, for example in enhancing distributed sensing [7], in-line filtering, as well as for suppressing undesired modes in large-core high-power fiber lasers [3, 5].

Certain designs of photonic crystal fibers can significantly suppress leakage without a bandgap [29-31], but residual radiation loss remains. Fibers with separable permittivity profiles have been suggested [32] but cannot be realized as exact values of permittivity need to be achieved. One way to realize BICs is by introducing large index modulations periodically along the propagation direction [18, 21, 22, 24-28, 33-47]; longitudinal periodic Bragg gratings in fibers can be created by writing interference fringes onto a photosensitive core or by femtosecond laser writing [48-55], but the available grating index contrast $\Delta_n$ is much smaller (of the order of $10^{-6} \sim 10^{-2}$) [48, 50, 53] than those that are known to support BICs. Therefore, despite much interest, it remains an open question whether BICs can exist in realistic optical fiber structures.

Here we use rigorous coupled-wave analysis (RCWA) to show that BICs do exist in realistic fiber Bragg gratings, that they do not require fine structural tuning or index tuning, and that they persist even in the limit of vanishingly small grating index contrasts. In addition to bona fide BICs with zero radiation loss, we also find quasi-BICs where the radiation loss can be suppressed by orders of magnitude. An analytical two-wave coupling analysis, appropriate for low-contrast gratings, reveals the conceptual origin of the BICs as the coupling between a high-quality-factor band and a low-quality-factor band. Our results indicate that BICs readily exist in a wide range of fiber Bragg grating structures.

## 2. RCWA FORMALISM

We consider a step-index fiber with periodic index modulations in the core [Fig. 1(a)]. The core has radius r and is surrounded by a cladding with relative permittivity $\varepsilon_d$. For concreteness, we consider the index modulation to consist of alternating dielectric layers with thicknesses d and $a - d$ at relative permittivities $\varepsilon_2$ and $\varepsilon_1$; our formalism is general and also treats other periodic profiles. The permittivity contrast is defined as $\Delta_\varepsilon = (\varepsilon_2 - \varepsilon_1)/\varepsilon_1$, and the index contrast is $\Delta_n \approx \Delta_\varepsilon/2$. The cladding's outer surface is not considered since the field is exponentially small there for the BICs of interest. Given the cylindrical symmetry, the fiber modes have $\exp(im\phi)$ angular dependence with distinct angular momentum indices $m$, where $\phi$ is the azimuthal angle. We start by considering fiber modes with $m = 0$, for which the TE ($\boldsymbol{H} = H_z\hat{z} + H_\rho\hat{\rho}$, $\boldsymbol{E} = E_\phi\hat{\phi}$) and TM ($\boldsymbol{E} = E_z\hat{z} + E_\rho\hat{\rho}$, $\boldsymbol{H} = H_\phi\hat{\phi}$) polarizations decouple[6], and each mode couples only to radiation channels in one polarization. We label these pure-polarization modes as $TE_n$ and $TM_n$ where n is the radial mode index. The hybrid-polarization modes with $m \neq 0$ will be considered later in section 5.

The low index contrast and high quality factor of the leaky fiber modes post a challenge for the often-used finite-difference or finite-element numerical methods. To accurately describe the leaky fiber modes, we employ the Fourier modal method, also called rigorous coupled-wave analysis [56, 57] (RCWA), which handles the radiating fields analytically. Here we briefly summarize the RCWA formalism and our implementation. Fields of the $TE_n$ and $TM_n$ modes inside and outside the core (where the relative permittivity $\varepsilon$ is a function of z only) satisfy the wave equation

TE: $(\nabla^2 + k_0^2 \varepsilon) H_z(\rho, z) = 0$, TM: $(\nabla^2 + k_0^2 \varepsilon + \frac{\partial}{\partial z} \frac{\varepsilon'}{\varepsilon}) E_z = 0$, (1)

where $k_0 = \omega/c$, $c$ is the vacuum speed of light, and $\omega$ is the frequency which may be complex-valued ($\omega = \omega_r + i\omega_i$) with the imaginary part representing radiation loss. The lifetime of a leaky mode is quantified by its quality factor, $Q = -\omega_r/(2\omega_i)$. Inside the core ($\rho < r$), $\varepsilon$ is periodic with $\varepsilon(z + a) = \varepsilon(z)$. Outside the core ($\rho > r$), $\varepsilon = \varepsilon_d$. A TE fiber mode with propagation constant $k_z$ is a solution of Eq. (1) with outgoing boundary condition at $\rho \to \infty$, which can be written as

$$H_z(\rho, z) = \begin{cases} e^{ik_z z} \sum_p C_p u_p(z) \frac{J_0(\gamma_p \rho)}{J_0(\gamma_p r)}, & \rho < r, \\ e^{ik_z z} \sum_p T_{p,TE} e^{i(\frac{2\pi p}{a})z} \frac{H_0(\kappa_p \rho)}{H_0(\kappa_p r)}, & \rho > r, \end{cases}$$ (2)

where $u_p(z)$ are a set of local modes in the core satisfying

$$\left[ (\frac{\partial}{\partial z} + ik_z)^2 + k_0^2 \varepsilon(z) \right] u_p(z) = \gamma_p^2 u_p(z),$$ (3)

with $\gamma_p$ being the radial propagation constant of the local mode inside the core, and $\kappa_p = \sqrt{\varepsilon_d k_0^2 - (k_z + 2\pi p/a)^2}$ being the radial propagation constant outside. $J_0$ and $H_0$ are the zeroth Bessel function and Hankel function of the first kind. The coefficients $C_p$ and $T_{p,TE}$ determine the interior and exterior (both evanescent and radiating) fields. $p = 0, \pm 1, \pm 2, \ldots$ is the index of the local modes, labeled by the primary Fourier component of $u_p$. Given $k_z$, we numerically solve Eq. (3) in the Fourier basis for $\mu_p$ and $\beta_p^2$, and then impose continuity of $H_z$ and $E_\phi$ at $\rho = r$ via Eq. (2) to obtain the complex frequency $\omega = \omega_r + i\omega_i$ of the fiber mode and its field profile. In the following, we label the resulting TE fiber modes as $TE_n^{(P)}$, where P is the main Fourier component and $n$ is the radial index. The same method works for TM modes by solving for $E_z$ in Eq. (1). With this method, we obtain an exact solution of the vectorial Maxwell's equations, with the only approximation being a truncation of terms in the Fourier basis. This method remains robust when the index contrast is small. In fact, since the different terms in the summation are coupled through the periodic index modulation, smaller contrast means weaker coupling and fewer terms to keep; we will take advantage of this property to develop a two-term approximation in section 4.

To validate our RCWA implementation, we consider a grating with high index contrast, for which the radiation loss is significant and conventional numerical methods are adequate; such case is similar to the chain-of-disks structure recently studied in Refs. [28, 46, 47]. Supplementary Fig. S1 compares results from our RCWA implementation with those from conventional finite-element frequency-domain simulations for the case $\varepsilon_2 = 2.16$, $\varepsilon_1 = \varepsilon_d = 1$; the perfect agreement validates our calculations.

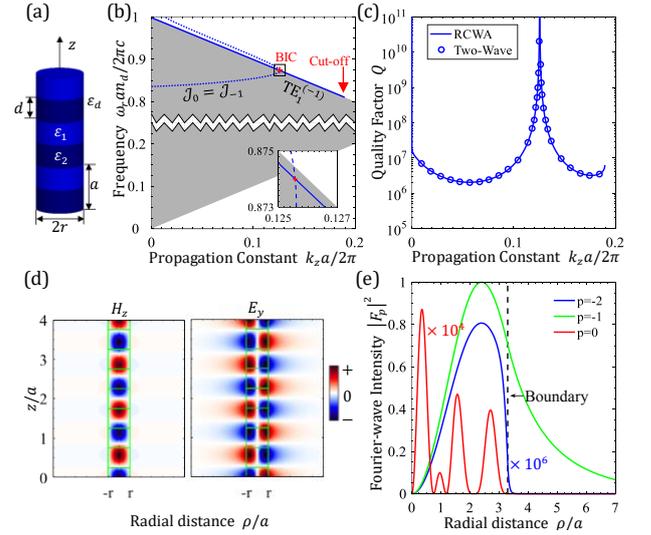

**Fig. 1.** Bound state in the continuum (BIC) in fiber Bragg gratings. **(a)** Schematic of a fiber Bragg grating considered here. The fiber core is cylindrically symmetric, periodic in the z direction with periodicity $a$, and surrounded by a cladding with relative permittivity $\varepsilon_d$. **(b)** Band structure and **(c)** quality factor of the $TE_1^{(-1)}$ mode. The cladding has index $n_d = 1.444$. The core has index $\sqrt{\varepsilon_1} = 1.455$, radius $r = 3.3a$, with permittivity contrast $\Delta_\varepsilon = (\varepsilon_1 - \varepsilon_2)/\varepsilon_1 = 10^{-2}$; $\Delta_n = 5 \times 10^{-3}$. The quality factor diverges at $k_z \approx 0.126 \cdot (2\pi/a)$, where the leaky mode turns into a BIC. The inset of (b) shows a zoom-in close to the BIC. The blue dotted curve is the solution of $\mathcal{J}_0 = \mathcal{J}_{-1}$ described in section 4; its intersection with the $TE_1^{(-1)}$ band is the location of the BIC, marked with a red plus. The solid curve and the circles in (c) are from full RCWA solution (including 11 Fourier terms) and the two-wave approximation, respectively. **(d)** Mode profile and **(e)** intensities of the main Fourier components of the BIC. The $|F_{-2}|^2$ and $|F_0|^2$ curves are multiplied by $10^6$ and $10^4$ to be more visible.

## 3. BOUND STATES IN THE CONTINUUM IN FIBER BRAGG GRATINGS

We now consider a realistic fiber Bragg grating, with a small grating contrast of $\Delta_\varepsilon = 10^{-2}$ ($\Delta_n = 5 \times 10^{-3}$). The fiber parameters are chosen to be comparable to those of a commercial fiber (Fibercore SM1500SC). Here we examine the leaky $TE_1^{(-1)}$ mode, whose dispersion calculated from RCWA is shown in Fig. 1(b). At frequencies above the light line, $\omega_r > |k_z|c/\sqrt{\varepsilon_d}$, the fiber mode couples to the continuum of free-space modes in the cladding and is generally leaky with a finite quality factor. However, we find that its quality factor diverges to infinity at a discrete propagation constant $k_z \approx 0.126 \times 2\pi/a$, where it becomes a lossless BIC [Fig. 1(c)]. The field profile of the BIC [Fig. 1(d)] exhibits exponential decay outside the fiber core, with no outward propagation.

The disappearance of radiation can be quantified by the Fourier components $F_p(\rho)$ of the field profile, defined via $E_\Phi(\rho, z) = e^{ik_z z} \sum_p [F_p(\rho) \exp(i2\pi pz/a)]$ and shown in Fig. 1(e). The zeroth Fourier component $F_0$ is the one that carries outgoing radiation in the cladding. For this BIC, we observe that $F_0$ is nonzero inside the fiber core, indicating there are interior fields that cannot be confined by total internal reflection. However, $F_0$ vanishes outside the core so does not carry outgoing radiation. This disappearance of radiation results from the destructive interference among radiation from multiple local modes $\mu_p$ in the rod.

We note that BICs may also arise through separability due to symmetry [15] as can be seen at $k_z = 0$ in Fig. 1(c); more details are given in supplementary Fig. S2. For such BICs, $F_0$ is zero both inside and outside the fiber core because of symmetry. These symmetry-protected BICs have zero group velocity and do not propagate along the fiber.

## 4. TWO-WAVE COUPLING ANALYSIS

The low grating contrast suggests that we can understand the leaky resonance and the BIC based on modes of a homogeneous fiber without grating. The dispersion of the TE modes for a homogeneous fiber is given by [6]

$$\frac{J_0'(\gamma_{0,p}r)}{\gamma_{0,p}J_0(\gamma_{0,p}r)} = \frac{H_0'(\kappa_p r)}{\kappa_p H_0(\kappa_p r)}, \quad (4)$$

where $\gamma_{0,p} = \sqrt{k_0^2 \varepsilon_c - (k_z + 2\pi p/a)^2}$, and $\varepsilon_c$ is the relative permittivity of the core. The solutions of Eq. (4) are plotted in Fig. 2(a,b). Here the Fourier index $p$ of each band $\text{TE}_n^{(p)}$ corresponds to band folding in the reduced-zone scheme as we impose an artificial periodicity $a$. In this homogeneous case, each band contains a single Fourier component with $\mu_p(z) = e^{i(2\pi p/a)z}$, and there is no coupling between bands. Fiber modes with frequency above the unfolded light line ($\omega_r > |k_z + 2\pi p/a|c/\sqrt{\varepsilon_d}$) are leaky and radiate strongly, as can be seen from their Q factors [Fig. 2(b)]. Those below the unfolded light line are index guided and do not radiate. In Fig. 2 we illustrate using a fiber without cladding ($\varepsilon_d = 1$) and with periodicity $a = r$ so that the relevant fiber modes are easier to see in the band structure.

The periodic index modulation couples the different Fourier components. Modes above the light line that were guided in the homogeneous fiber can now radiate by coupling to the leaky modes with $p = 0$, analogous to the guided resonances in photonic crystal slabs[58]. Because of the low index modulation, the dominant Fourier component $F_{-1}$ of the $\text{TE}_1^{(-1)}$ band only couples appreciably to the neighboring components $F_0$ and $F_{-2}$; the other components are orders of magnitude smaller. Among these three central components, $F_{-2}$ does not contribute to radiation and only makes up a small portion of the mode profile. Away from $k_z = 0$ and from the higher order modes, it is therefore a good approximation to keep only the $F_0$ and $F_{-1}$ components in RCWA. As shown in the Supplementary section II, this two-wave coupling analysis leads to the dispersion equation of $\text{TE}_n^{(0)}$ and $\text{TE}_{n'}^{(-1)}$ modes as

$(\mathcal{J}_0 - \mathcal{H}_0)(\mathcal{J}_{-1} - \mathcal{H}_{-1}) + (\mathcal{J}_0 - \mathcal{H}_{-1})(\mathcal{J}_{-1} - \mathcal{H}_0)\sigma^2 = 0, \quad (5)$

where $\mathcal{J}_p \triangleq J_0'(\gamma_p r)/[\gamma_p J_0(\gamma_p r)]$, $\mathcal{H}_p \triangleq H_0'(\kappa_p r)/\kappa_p H_0(\kappa_p r)$, with $p = -1, 0$. When $\sigma \to 0$, Eq. (5) reduces to Eq. (4). This dimensionless coefficient $\sigma = k_0^2 \varepsilon_\Delta/(\gamma_{-1}^2 - \gamma_0^2)$ couples the two bands $\text{TE}_n^{(0)}$ and $\text{TE}_{n'}^{(-1)}$, with $\varepsilon_\Delta$ being the first-order Fourier coefficient of the core index profile $\varepsilon(z)$. For the alternating-index grating here, $\varepsilon_\Delta = \varepsilon_1 \Delta(d/a)\text{sinc}(d/a)$. Away from the Brillouin zone edge, $\gamma_0^2 \approx \gamma_{0,0}^2 - k_0^4 \varepsilon_\Delta^2/(\gamma_{0,-1}^2 - \gamma_{0,0}^2)$, $\gamma_{-1}^2 \approx \gamma_{0,-1}^2 + k_0^4 \varepsilon_\Delta^2/(\gamma_{0,-1}^2 - \gamma_{0,0}^2)$. Fig. 2(c,d) shows the $\text{TE}_1^{(-1)}$ solution of Eq. (5) for three different grating permittivity contrasts ($\Delta_\varepsilon = 10^{-2}, 10^{-3}, 10^{-4}$) in blue, green, and orange solid curves, which agree quantitatively with the full RCWA solutions (shown in triangles) except at $k_z = 0$ where three Fourier components are needed. The corresponding solution for the example in Fig. 1 is shown as blue circles in Fig. 1(c).

The preceding analysis shows that to explain the BICs of interest in low-contrast fiber Bragg gratings, it is sufficient to consider the coupling between two bands, $\text{TE}_2^{(0)}$ and $\text{TE}_1^{(-1)}$. This may be compared to the coupled-resonances model of Friedrich and Wintgen [15, 17, 59-63]. In the Friedrich–Wintgen model, two resonances radiate into the same channel, and the overlap of radiation leads to the BIC. Here, one of the two bands $\text{TE}_1^{(-1)}$ does not radiate in the absence of the grating [Fig. 2(a,b)]. The Bragg grating that couples the two bands is responsible for both its radiation and the subsequent suppression of its radiation at the BIC.

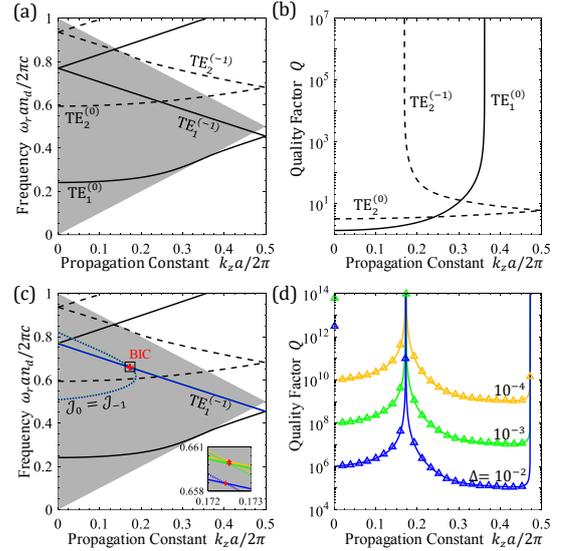

**Fig. 2.** Illustration of the two-wave coupling analysis. (**a-b**) Modes of a homogeneous fiber with dielectric $\varepsilon_1 = \varepsilon_2 \equiv \varepsilon_c = 2.12$ in $\varepsilon_d = 1$. (**a**) Band structure of the TE modes with $m = 0$. By artificially setting a period of $a = r$ along the $z$ direction, the dispersion curves are folded into the Brillouin zone. Solid and dashed curves are $\text{TE}_1^{(p)}$ and $\text{TE}_2^{(p)}$ modes with different Fourier indices $p$. The shaded area marks the radiation continuum with one leaky channel. (**b**) Quality factor $Q$ of the leaky fiber modes. The $\text{TE}_1^{(-1)}$ band is not shown as it has infinite $Q$ in the homogeneous fiber. (**c-d**) Dispersion curves in the presence of a low-contrast grating, which couples the $\text{TE}_2^{(0)}$ and $\text{TE}_1^{(-1)}$ bands. (**c**) Band structure and (**d**) quality factors of the $\text{TE}_1^{(-1)}$ modes. The blue, green and orange solid curves are two-wave approximations with the grating permittivity contrast $\Delta_\varepsilon$ being $10^{-2}$, $10^{-3}$ and $10^{-4}$, respectively, and the colored triangles are full RCWA solutions including 11 Fourier orders. The inset of (**c**) is a zoom-in close to the BICs, and the dotted curves are the solutions of $\mathcal{J}_0 = \mathcal{J}_{-1}$.

The two-wave coupling analysis also yields a simple analytical prediction for the locations of the BICs. A BIC is the simultaneous solution of two equations (Supplementary section II):

$$\mathcal{J}_0 = \mathcal{J}_{-1}, \quad \mathcal{J}_{-1} = \mathcal{H}_{-1}. \quad (6)$$

The first equation guarantees no radiation ($T_0 = 0$), while the second one is essentially the homogeneous-fiber dispersion [Eq. (4) except for the minor difference between $\gamma_{-1}$ and $\gamma_{0,-1}$ that is negligible at low contrast]. Both equations have solutions at real-valued frequencies. Therefore, the two sets of solution curves can intersect at real-valued frequencies corresponding to the BICs. The solution of $\mathcal{J}_0 = \mathcal{J}_{-1}$ is plotted in blue dashed curves in Fig. 1(b) and Fig. 2(c). Indeed, its intersection with the homogeneous-fiber dispersion gives the BICs marked by red pluses. Being at the intersection of two curves, the BIC is robust under small parameter changes of the system (which will only shift the curves), consistent with the topological interpretation of BICs [24, 26, 38, 41, 42]. The generic topological protection does not tell us whether a BIC can persist in the vanishing-contrast limit, since a BIC can disappear by annihilating with another BIC or by moving underneath the light line. However, here we know that the BICs of interest persist in such limit: for vanishingly small index contrasts ($\Delta_\varepsilon \to 0, \gamma_p \to \gamma_{0,p}$), the solutions of Eq. (6) approach a set of fixed points, as can be seen in Fig. 2(c). Therefore we now have an analytic proof that for small index contrasts, the BICs approach fixed locations in the band structure and do not vanish.

## 5. OTHER MODES AND QUASI-BICS

Having established the existence of BICs in the TE modes, we now extend our analysis to the other fiber modes. For the TM modes with $m = 0$ (for which $\boldsymbol{E} = E_z\hat{z} + E_\rho\hat{\rho}$, $\boldsymbol{H} = H_\phi\hat{\phi}$), $E_z$ in the cladding ($\rho > r$) is expressed as $E_z = e^{ik_z z}\sum_p T_{p,TM} e^{i(2\pi p/a)z} H_0(\kappa_p\rho)/H_0(\kappa_p r)$, similar to $H_z$ in Eq. (2); the rest is the same as described in section 2. When the angular momentum index $m$ is nonzero, the two polarizations are coupled, and there are two sets of radiation channels characterized by $T_{p,TM}$ and $T_{p,TE}$. For $HE_{11}^{(-1)}$ here, the relative radiation strengths are $S_{TE} = \mu_0|T_{0,TE}|^2/(\varepsilon_0 n_d^2)$ and $S_{TM} = |T_{0,TM}|^2$ (see Supplementary section III). The quality factors of the first few leaky fiber modes, including $TE_1^{(-1)}$, $TM_1^{(-1)}$, $HE_{21}^{(-1)}$, and the fundamental mode $HE_{11}^{(-1)}$, are shown in Fig. 3(a) for the structure considered in Fig. 1. Note that even though $TE_1^{(-1)}$, $TM_1^{(-1)}$, and $HE_{21}^{(-1)}$ are degenerate in the absence of the grating (together they make up the LP11 group [6]), they take on different radiation losses when the grating is introduced. In the $TM_1^{(-1)}$ band, two BICs can be readily identified; the mechanism is the same as the BICs in $TE_1^{(-1)}$ since there is only one radiation channel. The sharp variation of the quality factor near $k_z = 0$ (wavelength $\lambda \approx an_d$) is further detailed in supplementary Fig. S2.

The hybrid-polarization modes $HE_{11}^{(-1)}$ and $HE_{21}^{(-1)}$ couple to two radiation channels (in TE and TM polarizations). As shown in Fig. 3(b), each of the two radiation powers $S_{TE}$ and $S_{TM}$ crosses zero at discrete frequencies and propagation constants. Generically they vanish at different points, so there is no bona fide BIC in these hybrid-polarization modes. However, since $T_{0,TE}$ is the dominant radiation channel, the quality factor reaches local maxima near the zeros of $T_{0,TE}$, forming quasi-BICs where the quality factor is enhanced by almost two orders of magnitude. Despite the finite quality factor, such quasi-BICs can also be very useful for applications [15, 29-31, 63-66].

## 6. CONCLUSION

In conclusion, we have shown that BICs exist in fiber Bragg gratings with low index modulation contrasts. The BICs arise from the coupling between a high-Q band and a low-Q band, and they persist in the limit of vanishingly small index contrasts. These are very encouraging results, as they indicate that BICs may be realized in practical optical fibers. Such fiber BICs have wavelength and propagation constant selectivity, and can have many uses in fiber-based optical devices including fiber filters, sensors, and lasers.

**Funding.** X.G. and H.C. were supported by the National Natural Science Foundation of China under Grants No. 61625502, No. 61574127, No. 61601408, No. 61775193 and No. 11704332, the ZJNSF under Grant No. LY17F010008, the Top-Notch Young Talents Program of China, the Fundamental Research Funds for the Central Universities under Grant No. 2017XZZX008-06, and the Innovation Joint Research Center for Cyber-Physical-Society System. C.W.H. was supported by the National Science Foundation through grant no. DMR-1307632. B.Z. was supported by the Air Force Office of Scientific Research Young Investigator Program under award number FA9550-18-1-0133, the Charles E. Kaufman Foundation, a supporting organization of the Pittsburgh Foundation, and NSF through the University of Pennsylvania Materials Research Science and Engineering Center DMR-1720530. M.S. was supported by the MRSEC Program of the National Science Foundation under award number DMR-1419807 and the Army Research Office under Cooperative Agreement Number W911NF-18-2-0048.

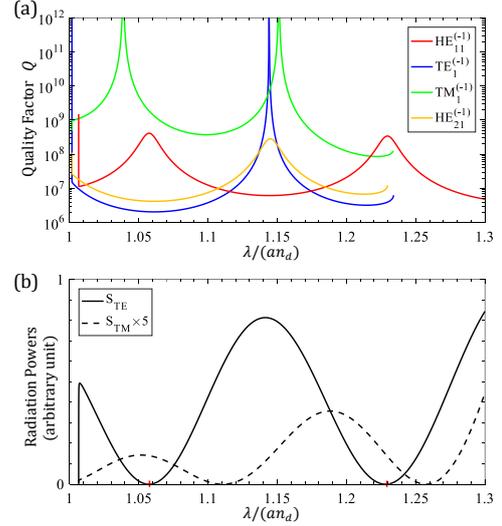

**Fig. 3 (a)** RCWA result of the quality factors of the first few fiber modes for the fiber Bragg grating considered in Fig. 1, as a function of the normalized wavelength $\lambda = 2\pi c/\omega_r$. **(b)** The two radiation strengths of the $HE_{11}^{(-1)}$ mode. Red crosses indicate the peaks of $HE_{11}^{(-1)}$'s quality factor. The TM-polarized radiation strength $S_{TM}$ is multiplied by 5 to be more visible.

## References

1. Agrawal, G.P., *Nonlinear fiber optics*. 2007: Academic press.
2. Fermann, M.E. and I. Hartl, *Ultrafast fibre lasers.* Nat Photon, 2013. **7**(11): p. 868-874.
3. Jauregui, C., J. Limpert, and A. Tunnermann, *High-power fibre lasers.* Nat Photon, 2013. **7**(11): p. 861-867.
4. Xu, C. and F.W. Wise, *Recent advances in fibre lasers for nonlinear microscopy.* Nat Photon, 2013. **7**(11): p. 875-882.
5. Zervas, M.N. and C.A. Codemard, *High Power Fiber Lasers: A Review.* IEEE Journal of Selected Topics in Quantum Electronics, 2014. **20**(5): p. 219-241.
6. Agrawal, G.P., *Fiber-optic communication systems*. Vol. 222. 2012: John Wiley & Sons.
7. Richardson, D.J., J.M. Fini, and L.E. Nelson, *Space-division multiplexing in optical fibres.* Nat Photon, 2013. **7**(5): p. 354-362.
8. Lee, B., *Review of the present status of optical fiber sensors.* Optical fiber technology, 2003. **9**(2): p. 57-79.
9. Abouraddy, A.F., et al., *Towards multimaterial multifunctional fibres that see, hear, sense and communicate.* Nat Mater, 2007. **6**(5): p. 336-347.
10. Flusberg, B.A., et al., *Fiber-optic fluorescence imaging.* Nat

Meth, 2005. **2**(12): p. 941-950.
11. Temelkuran, B., et al., *Wavelength-scalable hollow optical fibres with large photonic bandgaps for CO2 laser transmission.* Nature, 2002. **420**(6916): p. 650-653.
12. Knight, J.C., *Photonic crystal fibres.* nature, 2003. **424**(6950): p. 847-851.
13. Russell, P., *Photonic Crystal Fibers.* Science, 2003. **299**(5605): p. 358-362.
14. Joannopoulos, J.D., et al., *Photonic crystals: molding the flow of light.* 2011: Princeton university press.
15. Hsu, C.W., et al., *Bound states in the continuum.* Nature Reviews Materials, 2016. **1**: p. 16048.
16. Von Neumann, J. and E. Wigner, *Uber merkwürdige diskrete Eigenwerte. Uber das Verhalten von Eigenwerten bei adiabatischen Prozessen.* Physikalische Zeitschrift, 1929. **30**: p. 467-470.
17. Friedrich, H. and D. Wintgen, *Interfering resonances and bound states in the continuum.* Physical Review A, 1985. **32**(6): p. 3231.
18. Hsu, C.W., et al., *Observation of trapped light within the radiation continuum.* Nature, 2013. **499**(7457): p. 188-91.
19. Corrielli, G., et al., *Observation of Surface States with Algebraic Localization.* Physical Review Letters, 2013. **111**(22): p. 220403.
20. Weimann, S., et al., *Compact Surface Fano States Embedded in the Continuum of Waveguide Arrays.* Physical Review Letters, 2013. **111**(24): p. 240403.
21. Gansch, R., et al., *Measurement of bound states in the continuum by a detector embedded in a photonic crystal.* Light: Science and Applications, 2016. **5**(9): p. e16147.
22. Kodigala, A., et al., *Lasing action from photonic bound states in continuum.* Nature, 2017. **541**(7636): p. 196-199.
23. Gomis-Bresco, J., D. Artigas, and L. Torner, *Anisotropy-induced photonic bound states in the continuum.* Nature Photonics, 2017. **11**(4): p. 232-236.
24. Zhang, Y., et al., *Observation of Polarization Vortices in Momentum Space.* Physical Review Letters, 2018. **120**(18): p. 186103.
25. Zhang, W., et al., *Extraordinary optical reflection resonances and bound states in the continuum from a periodic array of thin metal plates.* Optics Express, 2018. **26**(10): p. 13195-13204.
26. Doeleman, H.M., et al., *Experimental observation of a polarization vortex at an optical bound state in the continuum.* Nature Photonics, 2018. **12**(7): p. 397-401.
27. Bahari, B., et al., *Integrated and Steerable Vortex Lasers using Bound States in Continuum.* Bulletin of the American Physical Society, 2018.
28. Belyakov, M., et al., *Experimental observation of symmetry protected bound state in the continuum in a chain of dielectric disks.* arXiv preprint arXiv:1806.01932, 2018.
29. Couny, F., et al., *Generation and Photonic Guidance of Multi-Octave Optical-Frequency Combs.* Science, 2007. **318**(5853): p. 1118-1121.
30. Debord, B., et al., *Ultralow transmission loss in inhibited-coupling guiding hollow fibers.* Optica, 2017. **4**(2): p. 209-217.
31. Bulgakov, E.N. and A.F. Sadreev, *Fibers based on propagating bound states in the continuum.* arXiv preprint arXiv:1804.06626, 2018.
32. Birks, T.A., et al. *Strictly-bound modes of an idealised hollow-core fibre without a photonic bandgap.* in *36th European Conference and Exhibition on Optical Communication.* 2010.
33. Shipman, S.P. and S. Venakides, *Resonance and Bound States in Photonic Crystal Slabs.* SIAM Journal on Applied Mathematics, 2003. **64**(1): p. 322-342.
34. Marinica, D.C., A.G. Borisov, and S.V. Shabanov, *Bound States in the continuum in photonics.* Physical Review Letters, 2008. **100**(18): p. 183902.
35. Liu, V., M. Povinelli, and S. Fan, *Resonance-enhanced optical forces between coupled photonic crystal slabs.* Optics Express, 2009. **17**(24): p. 21897-21909.
36. Hsu, C.W., et al., *Bloch surface eigenstates within the radiation continuum.* Light Sci Appl, 2013. **2**: p. e84.
37. Yang, Y., et al., *Analytical perspective for bound states in the continuum in photonic crystal slabs.* Physical Review Letters, 2014. **113**(3): p. 037401.
38. Zhen, B., et al., *Topological Nature of Optical Bound States in the Continuum.* Physical Review Letters, 2014. **113**(25): p. 257401.
39. Bulgakov, E.N. and A.F. Sadreev, *Light trapping above the light cone in one-dimensional array of dielectric spheres.* Physical Review A, 2015. **92**(2).
40. Gao, X., et al., *Formation mechanism of guided resonances and bound states in the continuum in photonic crystal slabs.* Scientific Reports, 2016. **6**.
41. Bulgakov, E.N. and D.N. Maksimov, *Topological Bound States in the Continuum in Arrays of Dielectric Spheres.* Physical Review Letters, 2017. **118**(26): p. 267401.
42. Bulgakov, E.N. and D.N. Maksimov, *Bound states in the continuum and polarization singularities in periodic arrays of dielectric rods.* Physical Review A, 2017. **96**(6): p. 063833.
43. Xiao, Y.-X., et al., *Topological Subspace-Induced Bound State in the Continuum.* Physical Review Letters, 2017. **118**(16): p. 166803.
44. Yuan, L. and Y.Y. Lu, *Strong resonances on periodic arrays of cylinders and optical bistability with weak incident waves.* Physical Review A, 2017. **95**(2): p. 023834.
45. Yuan, L. and Y.Y. Lu, *Bound states in the continuum on periodic structures: perturbation theory and robustness.* Optics Letters, 2017. **42**(21): p. 4490-4493.
46. Bulgakov, E.N. and A.F. Sadreev, *Bound states in the continuum with high orbital angular momentum in a dielectric rod with periodically modulated permittivity.* Physical Review A, 2017. **96**(1): p. 013841.
47. Bulgakov, E.N. and A.F. Sadreev, *Scattering plane waves by a dielectric cylinder with periodically modulated permittivity at oblique incidence.* Physical Review A, 2018. **97**(6): p. 063856.
48. Philip St, J.R., A. Jean-Luc, and R. Laurence, *Fibre gratings.* Physics World, 1993. **6**(10): p. 41.
49. Hill, K.O. and G. Meltz, *Fiber Bragg grating technology fundamentals and overview.* Journal of lightwave technology, 1997. **15**(8): p. 1263-1276.
50. Othonos, A., *Fiber Bragg gratings.* Review of Scientific Instruments, 1997. **68**(12): p. 4309-4341.
51. Kashyap, R., *Chapter 2 - Photosensitivity and Photosensitization of Optical Fibers*, in *Fiber Bragg Gratings (Second Edition)*. 2010, Academic Press: Boston. p. 15-51.
52. Thomas, J., et al., *Femtosecond pulse written fiber gratings: a new avenue to integrated fiber technology.* Laser & Photonics

53. Homoelle, D., et al., *Infrared photosensitivity in silica glasses exposed to femtosecond laser pulses.* Optics Letters, 1999. **24**(18): p. 1311-1313.
54. Martinez, A., et al., *Direct writing of fibre Bragg gratings by femtosecond laser.* Electronics Letters, 2004. **40**(19): p. 1170-1172.
55. Sun, N.-H., et al. *Coupling Light into Fiber Using Second Order Fiber Bragg Gratings*. in *CLEO: 2013*. 2013. San Jose, California: Optical Society of America.
56. Liu, V. and S. Fan, *S 4: A free electromagnetic solver for layered periodic structures.* Computer Physics Communications, 2012. **183**(10): p. 2233-2244.
57. Armaroli, A., et al., *Three-dimensional analysis of cylindrical microresonators based on the aperiodic Fourier modal method.* JOSA A, 2008. **25**(3): p. 667-675.
58. Fan, S. and J. Joannopoulos, *Analysis of guided resonances in photonic crystal slabs.* Physical Review B, 2002. **65**(23): p. 235112.
59. Neukammer, J., et al., *Autoionization Inhibited by Internal Interferences.* Physical Review Letters, 1985. **55**(19): p. 1979-1982.
60. Sadreev, A.F., E.N. Bulgakov, and I. Rotter, *Bound states in the continuum in open quantum billiards with a variable shape.* Physical Review B, 2006. **73**(23): p. 235342.
61. Lepetit, T. and B. Kanté, *Controlling multipolar radiation with symmetries for electromagnetic bound states in the continuum.* Physical Review B, 2014. **90**(24): p. 241103.
62. Wiersig, J., *Formation of Long-Lived, Scarlike Modes near Avoided Resonance Crossings in Optical Microcavities.* Physical Review Letters, 2006. **97**(25): p. 253901.
63. Rybin, M.V., et al., *High-$Q$ Supercavity Modes in Subwavelength Dielectric Resonators.* Physical Review Letters, 2017. **119**(24): p. 243901.
64. Zhou, H., et al., *Perfect single-sided radiation and absorption without mirrors.* Optica, 2016. **3**(10): p. 1079-1086.
65. Taghizadeh, A. and I.-S. Chung, *Quasi bound states in the continuum with few unit cells of photonic crystal slab.* Applied Physics Letters, 2017. **111**(3): p. 031114.
66. Bogdanov, A., et al., *A direct link between Fano resonances and bound states in the continuum.* arXiv preprint arXiv:1805.09265, 2018.

# Supplementary:

## I. Fiber Bragg gratings with large index modulation

Fig. S1(**a-b**) shows the RCWA band structure and quality factors for a fiber Bragg grating with $\varepsilon_1 = 2.16$ and $\varepsilon_2 = 1$ (the permittivity contrast is $\Delta_\varepsilon = 54\%$) with no cladding ($\varepsilon_d = 1$). Most modes above the light line have strong radiation into the far field; a typical field profile is shown in Fig. S1(**c**) for a $\text{TE}_1^{(-1)}$ guided resonance at $k_z = 0.1 \cdot (2\pi/a)$. Fig. S1(**d**) shows the intensities of its $z$-directional Fourier coefficients. Meanwhile, at an isolated propagation constant $k_z \approx 0.25 \cdot (2\pi/a)$, the quality factor of the $\text{TE}_1^{(-1)}$ band diverges, and the leaky guided resonance turns into a BIC with no radiation. The corresponding mode profile of the BIC is shown in Fig. S1(**e**), with the intensities of the $z$-directional Fourier coefficients $F_p(\rho)$ shown in Fig. S1(**f**). Circles in Fig. S1(**a-b**) are finite-element frequency-domain simulation results using COMSOL, which agree with the RCWA results.

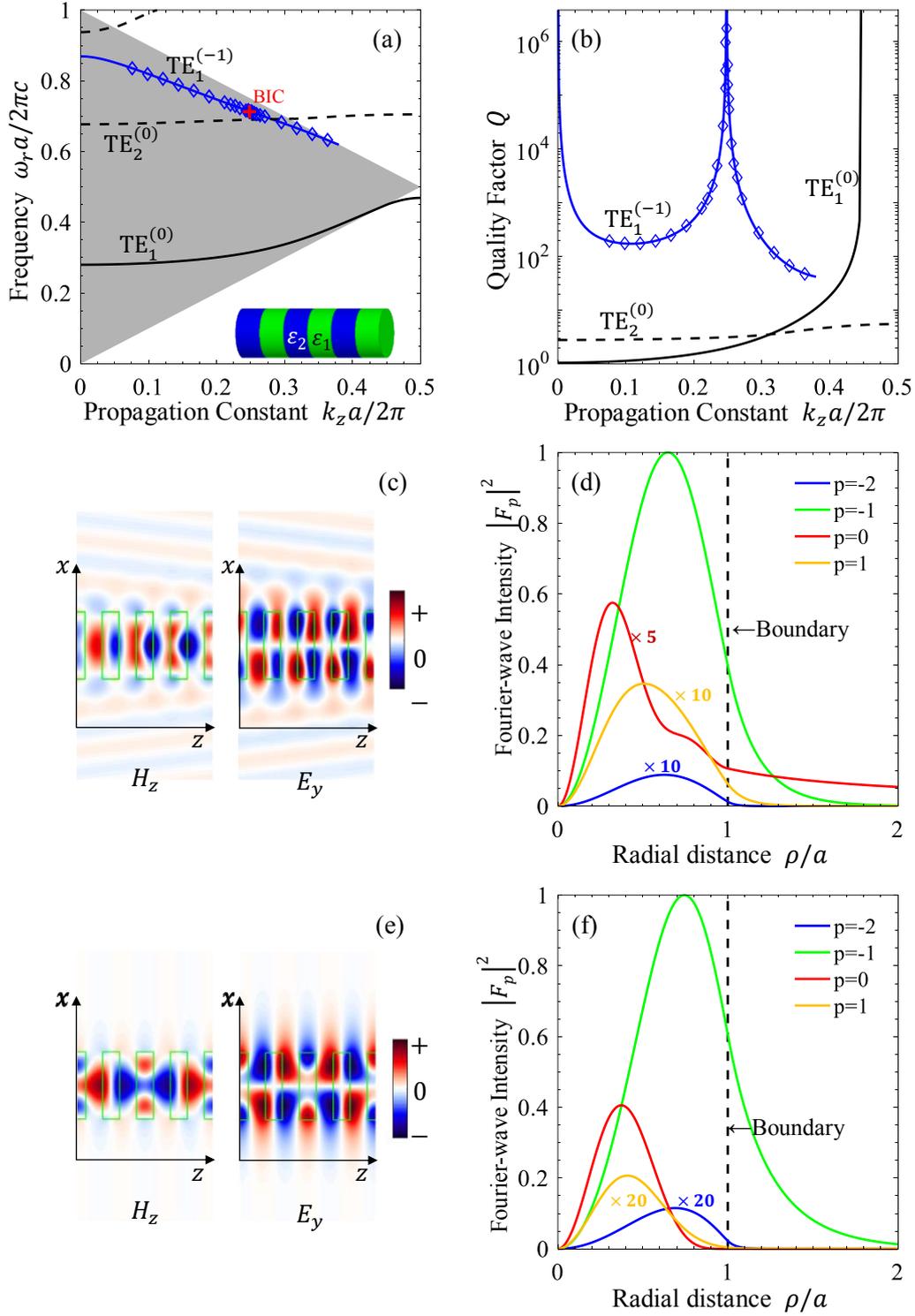

Fig. **S1**. RCWA solutions for a fiber Bragg grating with large index modulation. (**a**) Band structure and (**b**) quality factor of the TE modes for a system with $\varepsilon_2 = 2.16$, $\varepsilon_1 = \varepsilon_d = 1$, $r = a$, $d = 0.5a$ (schematic in the inset). The solid and dashed curves are RCWA solutions, and the blue diamonds are simulation results using COMSOL. Near propagation constant $k_z \approx 0.25 \cdot (2\pi/a)$ as marked by a red plus, the quality factor the $TE_1^{(-1)}$ band diverges, and the leaky guided resonance turns into a BIC that is not protected by symmetry. (**c**) Mode profile and (**d**) intensities of the main Fourier components of a guided resonance at $k_z = 0.1 \cdot (2\pi/a)$, $\omega \approx (0.82 - 0.0024i) \cdot (2\pi c/a)$. The $|F_{-2}|^2$, $|F_0|^2$ and $|F_1|^2$ curves are multiplied by appropriate factors to be more visible. (**e**) Mode profile and (**f**) intensities of the main Fourier components of the BIC. The $|F_{-2}|^2$, and $|F_1|^2$ curves are multiplied by a factor of 20 to be more visible. In (**c**) and (**d**), $H_z$ and $E_y$ are each normalized by their maximum amplitudes.

## II. Two-wave coupling analysis for TE modes

When truncated to two Fourier terms with $p = 0, -1$, local modes inside the core can be written as

$$u_p(z) = U_{0,p} + U_{-1,p} e^{-i(\frac{2\pi}{a})z}. \tag{S1}$$

Then, Eq. (3) in the main text becomes a two-by-two matrix equation

$$\mathbf{H}\mathbf{U}_p = \gamma_p^2 \mathbf{U}_p, \tag{S2}$$

with

$$\mathbf{U}_p = \begin{bmatrix} U_{0,p} \\ U_{-1,p} \end{bmatrix}, \quad \mathbf{H} = \begin{bmatrix} -k_z^2 & 0 \\ 0 & -(k_z - \frac{2\pi}{a})^2 \end{bmatrix} + k_0^2 \begin{bmatrix} \varepsilon_c & \varepsilon_\Delta \\ \varepsilon_\Delta & \varepsilon_c \end{bmatrix}, \quad \gamma_p^2 = \frac{\gamma_{0,-1}^2 + \gamma_{0,0}^2 \pm \sqrt{(\gamma_{0,-1}^2 - \gamma_0^2)^2 + 4k_0^4 \varepsilon_\Delta^2}}{2}.$$

$\varepsilon_c$ and $\varepsilon_\Delta$ are the zeroth and 1st (which equals to -1st) Fourier coefficients of $\varepsilon(z)$ in the core. If $k_z$ is not close to $\pi/a$, $\gamma_p^2$ is approximately

$$\gamma_0^2 \approx \gamma_{0,0}^2 - \frac{k_0^4 \varepsilon_\Delta^2}{\gamma_{0,-1}^2 - \gamma_{0,0}^2}, \quad \gamma_{-1}^2 \approx \gamma_{0,-1}^2 + \frac{k_0^4 \varepsilon_\Delta^2}{\gamma_{0,-1}^2 - \gamma_{0,0}^2}. \tag{S3}$$

The eigenvectors can be written as

$$\mathbf{P} \triangleq [\mathbf{U}_0, \mathbf{U}_{-1}] = \begin{bmatrix} 1 & \sigma \\ -\sigma & 1 \end{bmatrix}, \tag{S4}$$

where $\sigma = \frac{k_0^2 \varepsilon_\Delta}{\gamma_{0,-1}^2 - \gamma_{0,0}^2}$ couples the two local modes. To proceed, we need to determine the coefficients $C_p$ and $T_p$ in Eq. (2). For TE polarization, tangential electrical field is linked to $H_z$ as

$$k_0^2 \varepsilon E_\phi + \frac{\partial^2}{\partial z^2} E_\phi = -i\omega\mu_0 \frac{\partial H_z}{\partial \rho}. \tag{S5}$$

By substituting Eq. (2) into Eq. (S5), $E_\phi$ is derived in form of $E_\phi = \Phi_0(\rho) e^{i k_z z} + \Phi_{-1}(\rho) e^{i(k_z - 2\pi/a)z}$, where

$$\begin{bmatrix} \Phi_0 \\ \Phi_{-1} \end{bmatrix} = \mathbf{H}^{-1} \mathbf{P} \begin{bmatrix} \frac{\gamma_0 J_0'(\gamma_0 \rho)}{J_0(\gamma_0 r)} & 0 \\ 0 & \frac{\gamma_0 J_0'(\gamma_{-1} \rho)}{J_0(\gamma_{-1} r)} \end{bmatrix} \begin{bmatrix} C_0 \\ C_{-1} \end{bmatrix}, \quad \rho < r \tag{S6}$$

$$\begin{bmatrix} \Phi_0 \\ \Phi_{-1} \end{bmatrix} = \begin{bmatrix} \frac{H_0'(\kappa_0 \rho)}{\kappa_0 H_0(\kappa_0 r)} & 0 \\ 0 & \frac{H_0'(\kappa_{-1} \rho)}{\kappa_{-1} H_0(\kappa_{-1} r)} \end{bmatrix} \begin{bmatrix} T_0 \\ T_{-1} \end{bmatrix}, \quad \rho > r \tag{S7}$$

Eq. (S6) can be further simplified by an equivalent transformation of Eq. (S2): $\mathbf{H}^{-1}\mathbf{P} = \mathbf{P}\mathbf{B}^{-1}$, in which $\mathbf{B} = \text{diag}([\gamma_0^2, \gamma_{-1}^2])$. Imposing continuity of $H_z$ and $E_\phi$ at $\rho = r$, we get

$$\begin{bmatrix} 1 & \sigma \\ -\sigma & 1 \end{bmatrix} \begin{bmatrix} C_0 \\ C_{-1} \end{bmatrix} = \begin{bmatrix} T_0 \\ T_{-1} \end{bmatrix}, \tag{S8}$$

$$\begin{bmatrix} 1 & \sigma \\ -\sigma & 1 \end{bmatrix} \begin{bmatrix} \frac{J_0'(\gamma_0 r)}{\gamma_0 J_0(\gamma_0 r)} & 0 \\ 0 & \frac{J_0'(\gamma_{-1} r)}{\gamma_{-1} J_0(\gamma_{-1} r)} \end{bmatrix} \begin{bmatrix} C_0 \\ C_{-1} \end{bmatrix} = \begin{bmatrix} \frac{H_0'(\kappa_0 r)}{\kappa_0 H_0(\kappa_0 r)} & 0 \\ 0 & \frac{H_0'(\kappa_{-1} r)}{\kappa_{-1} H_0(\kappa_{-1} r)} \end{bmatrix} \begin{bmatrix} T_0 \\ T_{-1} \end{bmatrix}. \tag{S9}$$

The dispersion equation for the fiber modes (including both leaky guided resonances and BICs) is then given by combining Eq. (S8) with Eq. (S9) to yield

$$(\mathcal{J}_0 - \mathcal{H}_0)(\mathcal{J}_{-1} - \mathcal{H}_{-1}) + (\mathcal{J}_0 - \mathcal{H}_{-1})(\mathcal{J}_{-1} - \mathcal{H}_0)\sigma^2 = 0, \tag{S10}$$

where $\mathcal{J}_p \triangleq \frac{J_0'(\gamma_p r)}{\gamma_p J_0(\gamma_p r)}$, $\mathcal{H}_p \triangleq \frac{H_0'(\kappa_p r)}{\kappa_p H_0(\kappa_p r)}$. This is Eq. (5) in the main text.

For a BIC, there is no radiation field. By setting $T_0 = 0$, Eq. (S7) and Eq. (S8) produce two linear equations:

$$\begin{cases} C_0 + \sigma C_{-1} = 0 \\ \mathcal{J}_0 C_0 + \mathcal{J}_{-1} \sigma C_{-1} = 0 \end{cases} \tag{S11}$$

which combine to yield
$$\mathcal{J}_0 = \mathcal{J}_{-1}. \tag{S12}$$
A BIC is a joint solution of Eq. (S10) and Eq. (S12). Inserting Eq. (S12) into Eq. (S10) yields
$$\mathcal{J}_{-1} = \mathcal{H}_{-1}. \tag{S13}$$
Eq. (S12) and (S13) are Eq. (6) in the main text.

The same method works for TM modes by considering $E_z$ and $H_\phi$ which satisfy
$$(\nabla^2 + k_0^2 \varepsilon + \frac{\partial}{\partial z}\frac{\varepsilon'}{\varepsilon})E_z = 0,$$
and
$$k_0^2 H_\phi + \frac{\partial}{\partial z}\frac{1}{\varepsilon}\frac{\partial H_\phi}{\partial z} = i\omega\varepsilon_0 \frac{\partial E_z}{\partial \rho}.$$

# III. Hybrid modes in fiber Bragg Gratings

In general, an eigenmode in fiber Bragg gratings is expressed as follows:

$$H_z(\rho,\phi,z) = \begin{cases} e^{ik_z z + im\phi}\sum_p C_{p,TE} u_p(z)\frac{J_m(\gamma_p \rho)}{J_m(\gamma_p r)}, & \rho < r, \\ e^{ik_z z + im\phi}\sum_p T_{p,TE} e^{i(\frac{2\pi p}{a})z}\frac{H_m(\kappa_p \rho)}{H_m(\kappa_p r)}, & \rho > r, \end{cases} \tag{S14}$$

$$E_z(\rho,\phi,z) = \begin{cases} e^{ik_z z + im\phi}\sum_p C_{p,TM} v_p(z)\frac{J_m(\beta_p \rho)}{J_m(\gamma_p r)}, & \rho < r, \\ e^{ik_z z + im\phi}\sum_p T_{p,TM} e^{i(\frac{2\pi p}{a})z}\frac{H_m(\kappa_p \rho)}{H_m(\kappa_p r)}, & \rho > r, \end{cases} \tag{S15}$$

$m \in \mathbb{N}$ is angular momentum index. $u_p$ and $v_p$ form the bases of TE and TM waves in the one-dimensional periodic system, with corresponding eigenvalues $\gamma_p^2$ and $\beta_p^2$, respectively. The two bases are determined by

$$\left[\left(\frac{\partial}{\partial z} + ik_z\right)^2 + k_0^2 \varepsilon\right] u_p(z) = \gamma_p^2 u_p(z), \tag{S16}$$

and

$$\left[\left(\frac{\partial}{\partial z} + ik_z\right)^2 + k_0^2 \varepsilon + \left(\frac{\partial}{\partial z} + ik_z\right)\frac{1}{\varepsilon}\frac{d\varepsilon}{dz}\right] v_p(z) = \beta_p^2 v_p(z). \tag{S17}$$

$C_{p,TE}$, $C_{p,TM}$, $T_{p,TE}$ and $T_{p,TM}$ are to be solved through boundary condition at $\rho = r$. The boundary condition includes continuity of four field components: $H_z$, $E_z$, $E_\phi$ and $H_\phi$, among which the transversal fields are related to the longitude as follows:

$$k_0^2 \varepsilon E_\phi + \frac{\partial^2}{\partial z^2} E_\phi = -ik_0 \eta_0 \frac{\partial H_z}{\partial \rho} + \frac{1}{\rho}\frac{\partial^2 E_z}{\partial \phi \partial z}, \tag{S18}$$

$$k_0^2 H_\phi + \frac{\partial}{\partial z}\frac{1}{\varepsilon}\frac{\partial H_\phi}{\partial z} = ik_0 \eta_0^{-1} \frac{\partial E_z}{\partial \rho} + \frac{1}{\rho}\frac{\partial}{\partial z}\frac{1}{\varepsilon}\frac{\partial H_z}{\partial \phi}, \tag{S19}$$

where $\eta_0 = \sqrt{\mu_0/\varepsilon_0}$. By solving Eqs (S16-S17), one can get the transformation matrices from $\{u_p\}$ and $\{v_p\}$ to $\{\exp[(i2\pi p/a)z]\}$ (Fourier expansion). Eqs (S14-15) and Eqs (S18-19) can thus be written in Fourier series. In this manner, the boundary condition reduces to a group of space-independent linear equations with respect to $C_{p,TE}$, $C_{p,TM}$, $T_{p,TE}$ and $T_{p,TM}$. Finally, the modes (dispersion relation) are solved by requiring nontrivial solutions.

When $m = 0$, the $\partial/\partial\phi$ terms in Eq (S18) and Eq (S19) vanish. That means $E_\phi$ is independent of $E_z$ and $H_\phi$ is independent of $H_z$. Consequently, the linear equations are split into two decoupled groups. One contains $H_z$ and $E_\phi$ while the other contains $E_z$ and $H_\phi$. The two equation groups give solutions to TE and TM modes, separately.

If $m \neq 0$, transversal fields are always connected to both $H_z$ and $E_z$. The overall solution is a hybrid mode.

Furthermore, since the cladding is homogeneous, $E_z(\rho > r)$ and $H_z(\rho > r)$ has the same radial propagating constant $\kappa_p$ [Eqs (S14-15)]. For any subscript $p$, once $\text{Real}(\kappa_p) > 0$, there will always be two radiation channels denoted by $T_{p,TE}$ and $T_{p,TM}$. A perfect confinement of light in the continuum requires inhibition of two radiation channels, which is possible but generally requires tuning a structural parameter in addition to tuning $k_z$.

For $\text{HE}_{11}^{(-1)}$ mode, the radiation waves are the zeroth Fourier terms,

$$H_{z0} = T_{0,TE} e^{ik_z z + i\phi} \frac{H_1(\kappa_0 \rho)}{H_1(\kappa_0 r)},$$

$$E_{z0} = T_{0,TM} e^{ik_z z + i\phi} \frac{H_1(\kappa_0 \rho)}{H_1(\kappa_0 r)}.$$

To compare TE and TM radiations, we go to the far-field region where $\rho \to +\infty$. Setting $\varepsilon = \varepsilon_d = n_d^2$, from Eq (S18) and Eq (S19) we obtain the far fields as follows:

$$E_{\phi 0} = -ik_0 \eta_0 T_{0,TE} e^{ik_z z + i\phi} \frac{H_1'(\kappa_0 \rho)}{\kappa_0 H_1(\kappa_0 r)},$$

$$H_{\phi 0} = ik_0 \varepsilon_d \eta_0^{-1} T_{0,TM} e^{ik_z z + i\phi} \frac{H_1'(\kappa_0 \rho)}{\kappa_0 H_1(\kappa_0 r)}.$$

The radial energy flux is $S_{TE}\hat{\rho} = (E_{\phi 0}\hat{\phi}) \times (H_{z0}^*\hat{z}) = E_{\phi 0} H_{z0}^* \hat{\rho}$ for TE wave and $S_{TM}\hat{\rho} = (E_{z0}\hat{z}) \times (H_{\phi 0}^*\hat{\phi}) = -E_{z0} H_{\phi 0}^* \hat{\rho}$ for TM wave. The ratio of $S_{TE}$ over $S_{TM}$ is

$$\frac{S_{TE}}{S_{TM}} = \frac{E_{\phi 0} H_{z0}^*}{-E_{z0} H_{\phi 0}^*} = \left(\frac{\eta_d |T_{0,TE}|}{|T_{0,TM}|}\right)^2, \tag{S20}$$

where $\eta_d = \eta_0 / n_d$ is the wave impedance in the cladding.

Fig. S2 shows the quality factors of the four lowest-order fiber modes at $k_z \approx 0$, with $p = -1$ being the lower-frequency branch (*i.e.*, the one shown in Fig. 1 and Fig. 3 in the main text) and $p = +1$ being the higher-frequency branch. The $\text{TE}_1^{(-1)}$ and $\text{TM}_1^{(+1)}$ modes have symmetry-protected BICs with a diverging $Q$ as $k_z$ approaches zero. For the hybrid modes, the quality factors of $\text{HE}_{11}^{(-1)}$ and $\text{HE}_{21}^{(-1)}$ rise up but saturate to a finite value. Due to the small grating index contrast, these features exist at small $k_z$ within a very small frequency range, so they are hard to see in Fig. 1c and Fig. 3a in the main text.

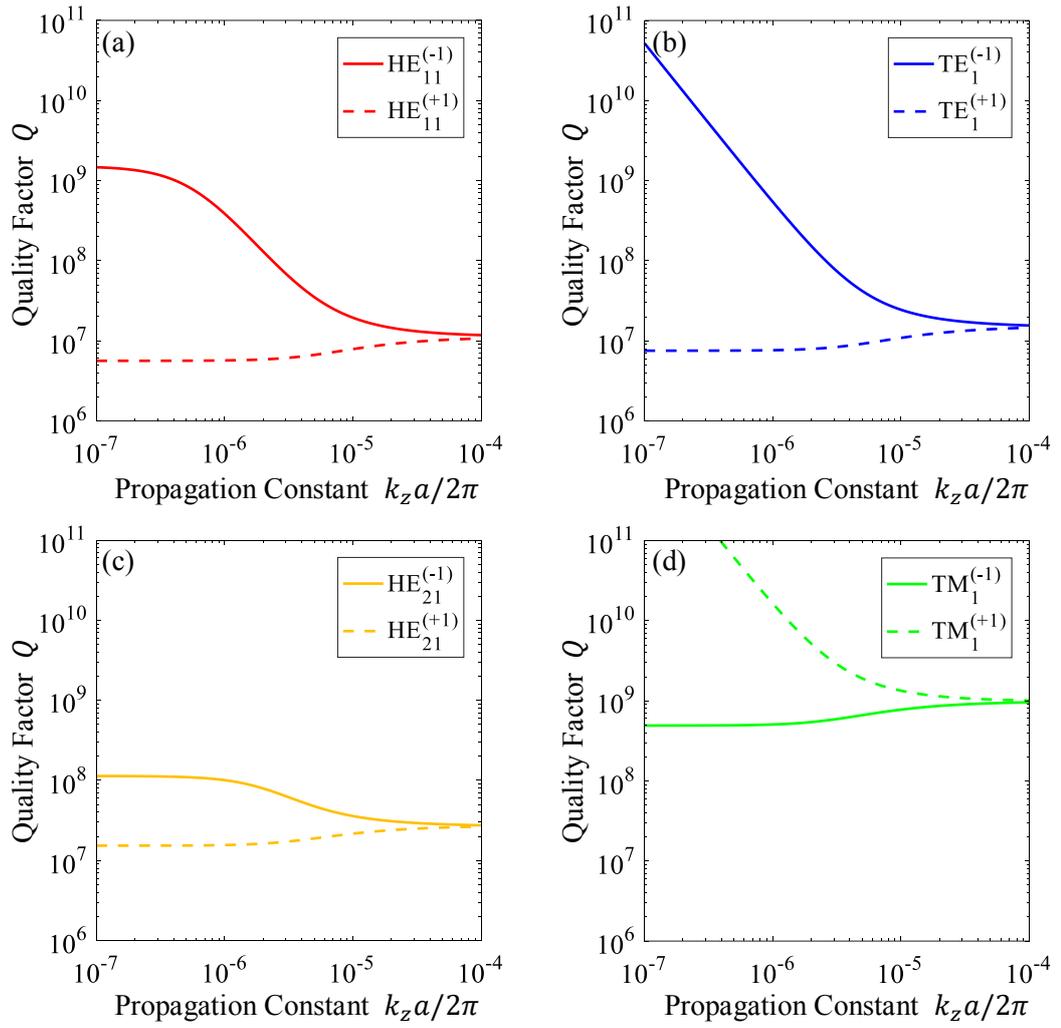

**Fig. S2.** Quality factors of the lowest four modes in Fig. 3(a), plotted within a small $k_z$ range. Solid curves represent $p = -1$ modes, and dashed curves represent $p = +1$ modes.